# Continuous Verification of Large Embedded Software using SMT-Based Bounded Model Checking

Lucas Cordeiro
University of Southampton
lcc08r@ecs.soton.ac.uk

Bernd Fischer
University of Southampton
b.fischer@ecs.soton.ac.uk

Joao Marques-Silva
University College Dublin
jpms@ucd.ie

## Abstract

*The complexity of software in embedded systems has increased significantly over the last years so that software verification now plays an important role in ensuring the overall product quality. In this context, SAT-based bounded model checking has been successfully applied to discover subtle errors, but for larger applications, it often suffers from the state space explosion problem. This paper describes a new approach called continuous verification to detect design errors as quickly as possible by looking at the Software Configuration Management (SCM) system and by combining dynamic and static verification to reduce the state space to be explored. We also give a set of encodings that provide accurate support for program verification and use different background theories in order to improve scalability and precision in a completely automatic way. A case study from the telecommunications domain shows that the proposed approach improves the error-detection capability and reduces the overall verification time by up to 50%.*

## 1  Introduction

Embedded computer systems are used in a wide range of sophisticated applications, such as mobile phones or set-top boxes providing internet connectivity. The functionality demanded for such systems has increased significantly and an increasing number of functions are implemented in software rather than hardware. As a consequence, the verification of the software design and the correctness of its implementation have become increasingly difficult.

Bounded model checking (BMC) has been successfully applied to verify embedded software and discovered subtle errors in real designs [3]. BMC generates verification conditions (VCs) that reflect the exact path in which a statement is executed, the context in which a given function is called, and the bit-accurate representation of the expressions. Proving the validity of these VCs remains the main performance bottleneck in verifying large embedded software, despite attempts to cope with increasing system complexity by applying SMT (Satisfiability Modulo Theories) solvers [1, 8, 12].

We address this bottleneck with a new concept called *continuous verification*. It aims to automatically detect design errors and integration problems as quickly as possible by exploiting information from the software configuration management (SCM) system, systematically focusing the verification effort on new or modified functions. We use equivalence checking to determine whether modified functions need to be re-verified formally and we use existing test cases to reduce the search space for the model checker, thus combining dynamic and static verification. We show that the continuous verification approach substantially reduces the verification time of large embedded software.

We also exploit the different background theories of SMT solvers and combine different theories and solvers, based on an analysis of the syntactic structure of a given ANSI-C program. We then demonstrate that this combination significantly improves the performance of software model checking for a wide range of embedded software benchmarks from the telecommunications domain. This is the first work that exploits the syntactic structure of an ANSI-C program in order to combine different encodings and SMT solvers for BMC of embedded software.

We describe how these new (compared to our previous work [8]) contributions are realized in ESBMC, the Efficient SMT-Based Bounded Model Checker. ESBMC extends CBMC [6] to support different theories and different SMT solvers and to make use of high-level information to simplify and reduce the unrolled formula size. Experimental results show that our approach scales significantly better than both the SAT-based and SMT-based versions of the CBMC model checker. In addition, the continuous verification approach and the combination of different encodings and solvers allow us to go deeper into the system (compared to software model checkers only) and explore more exhaustively the state space (compared to testing only). This hybrid solution is suitable for checking properties in large state spaces.

## 2 Continuous Verification

Our approach has its roots in the continuous integration (CI) practice described by Fowler [11]. CI relies on every developer to create and execute unit, functional and integration tests before committing their source code to a single source repository. It also assumes the existence of an automated unit test framework. The SCM is then used to perform the system build and test processes in a completely automatic way. In *continuous verification*, we use the same information (i.e., development history and test cases) in a different way to improve the coverage and substantially reduce the verification time throughout the development of a product or product line. We use SMT-based bounded model checking to verify for each system build that the entire system still satisfies all properties given as assertions by the designers, as well as a range of language-specific safety properties such as the absence of arithmetic under- and overflow, out-of-bounds array indexing, or nil-pointer dereferencing. Figure 1 shows the main elements and steps of the continuous verification approach; the gray boxes indicate core steps. Section 3 describes the software verification process in more detail.

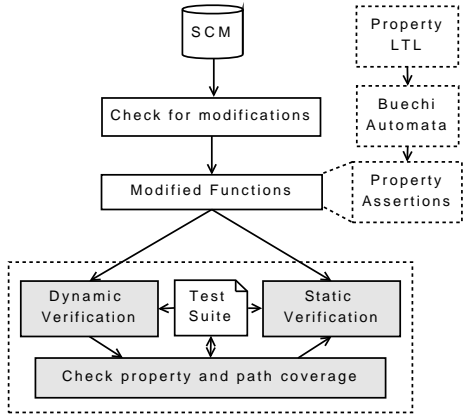

**Figure 1. Continuous Verification**

For large embedded software, the computational effort to re-verify the entire software from scratch is too high, and is largely wasted if, as is often the case, the changes are small [11]. For each system build, we thus consult the SCM to identify the functions and methods that have actually been modified and focus on these. We then use equivalence checking to determine whether they need to be re-verified formally: if we can prove that the old and new versions of a function are functionally equivalent, then we do not need to show for the new version any of the properties already shown for the old version. This reduces the immediate verification effort because proving the equivalence of two function versions is often less expensive than

```
unsigned signalInverter(int signal) {
  unsigned inverter;
  if(signal >= 0)
    inverter = signal;
  else
    inverter = -1*signal;
  return inverter;
}
```
(a)

```
unsigned signalInverter(int signal) {
  if(signal < 0)
    return -signal;
  else
    return signal;
}
```
(b)

**Figure 2. (a) Original function to invert the sign of signal. (b) Optimized version.**

re-verifying the function. However, and more importantly, it also reduces overall system verification efforts because it limits the propagation of changes through the system: if we can prove the two versions of the function computationally equivalent, then we do not need to re-verify any other function that depends it (unless that function has been changed as well). Of course, proving the equivalence of two functions is in general undecidable, due to unbounded memory usage [16], and the effort we spend in trying to do so might be wasted. However, in our experience, this does not occur very often, since a high degree of predictability is a desirable design characteristic in embedded software (i.e., dynamic memory allocations and recursion are discouraged).

As an example, consider the two versions of the *signalInverter* function shown in Figure 2. They were extracted from the embedded software of two releases of a medical device product. In order to prove the equivalence of these two ANSI-C functions, we compare their input-output relations. We thus first *(i)* remove from each function the variable declarations and return statements, *(ii)* convert the function bodies into single static assignment (SSA) form [17] (i.e., we introduce fresh variables by subscripting the original name such that every assignment has a unique left hand side), and *(iii)* conjoin all program statements. These operations produce two intermediate formulas $\alpha_1$ and $\alpha_2$ representing the functions' computations, as shown below.

$inverter_1 = signal_1$
$\wedge\ inverter_2 = -1 * signal_1$
$\wedge\ inverter_3 = (signal_1 \geq 0\ ?\ inverter_1 : inverter_2)$

$signal'_2 = (signal'_1 < 0\ ?\ -signal'_1 : signal'_1)$

For the actual equivalence check, we identify the input variables (i.e., $signal_1 = signal'_1$), and show using SMT-based bounded model checking that, given the representation of the function bodies, the output variables then also coincide:

$$(\alpha_1 \wedge \alpha_2 \wedge (signal_1 = signal'_1)) : (inverter_3 = signal'_2)$$

## 2.1 Specifying Temporal Properties with Büchi Automata

In addition to the language-specific safety properties, we can also show user-specified properties. These can be given directly as assertions in the code, using C's `assert` macro to state an assumption, or as formulas in linear-time temporal logic (LTL), which can track temporal properties of the software design. We translate the LTL formulas into Büchi automata using the Wring tool [21] and further into ANSI-C and merge them into the code. The resulting ANSI-C program then monitors the design's progress and watches out for violations of the specified properties (as described in Section 3).

As an example, we extract two properties from the specification of the medical device, and show how they can be modelled and used in the context of the continuous verification. The device, called a pulse oximeter [7], is responsible for measuring the oxygen saturation ($SpO_2$) and heart rate (HR) in the blood system using a non-invasive method. In particular, we verify (*a*) the data flow to compute the HR value that is provided by the pulse oximeter sensor hardware and (*b*) whether the user of the pulse oximeter is capable of adjusting the sample time of the embedded device. The properties (*a*) and (*b*) can be expressed using the following LTL pattern:

$$AG\,(p \rightarrow Fr) \qquad (1)$$

Here, $A$ ("for all paths"), $G$ ("always"), and $F$ ("eventually") are the LTL quantifiers, and $p$ and $r$ represent the required pre- and post-states. In the example, for the property (*a*), $p$ denotes the state that the buffer contains HR and $SpO_2$ raw data, while $r$ denotes the state that defines the respective HR value. Consequently, any state containing the HR and $SpO_2$ raw data in the buffer is eventually followed by a state representing the respective HR value.

A Büchi automaton is a finite automaton over infinite words. It differs from a standard finite automaton over finite words in the definition of accepting a word, which is based on passing through an accepting state infinitely often (rather than terminating in a final state) [5]. The Büchi automata we consider here work over computation traces, i.e., sequences of states of the program to be analyzed. These are abstracted by the predicates of interest (here $p$ and $r$). Hence the "words" can be represented by sequences of propositional expressions over the variables $p$ and $r$. Figure 3 shows the non-demininistic Büchi automaton that represents the LTL formula (1) and its corresponding ANSI-C monitor. The transition function $\delta$ is given by the following table.

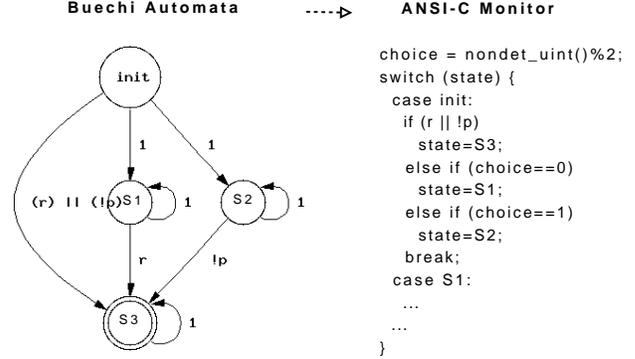

**Figure 3. Specifying Temporal Properties for Software.**

|      | 1          | $r \vee \neg p$ | $r$  | $\neg p$ |
|------|------------|-----------------|------|----------|
| init | $\{S1, S2\}$ | S3            | init | init     |
| S1   | S1         | S1              | S3   | S1       |
| S2   | S2         | S2              | S2   | S3       |
| S3   | S3         | S3              | S3   | S3       |

From the initial state, we can transition to S3 if $r \vee \neg p$ holds, stay in the initial state if either $r$, or $\neg p$ holds, or nondeterministically transition to either S1 or S2 if none of the three properties hold (denoted by 1). This automaton will accept all infinite words that represent computations in which each state in which $p$ holds will eventually be followed by a state in which $r$ holds. In order to model the nondeterministic transition of the Büchi-automata in the ANSI-C specification, we use a function *nondet_uint*() (which returns any number of type *unsigned int*) and then restricts its returning value to the domain $\{0, 1\}$. The property is then checked by using a "monitor" in such a way that the C program monitors the design's progress and watches out for a specific type of error up to the bound $k$. An assertion is then used to claim that an error is never encountered. In our example, we ensure that the buffer is not empty and contains the computed HR and $SpO_2$ values.

## 2.2 Generalizing Test Cases

After detecting new and/or modified functions, we use the existing unit test cases to reduce the state space to be explored by the model checker. In this phase, we first run the unit tests, keeping track of which inputs have already been used. We then guide the model checker to visit states that have not been visited previously. In addition, the test cases also help to reduce the state space to be explored in another way: by using the test stubs, we can break the global model (containing the entire program) into local models (containing only the functions under test) and generate on-demand the reachable states to be visited by the model checker, starting with the state described by the test case. We can so reduce the number of paths and variables to be considered during model checking.

As an example consider the three simple C functions shown in Figure 4 that were also extracted from the pulse oximeter embedded software, and one of the test cases shown in Figure 5. The functions implement a simple circular buffer using a FIFO (First In, First Out) policy. The test case checks whether messages are correctly added to and removed from the circular buffer using the FIFO policy. Other test cases check for buffer underflow and/or overflow and whether the elements are lost before reading them from the buffer.

```
static char    buffer[BUFFER_MAX];
void initLog(int max) {
   buffer_size = max;
   first = next = 0;
}

int removeLogElem(void) {
   first++;
   return buffer[first-1];
}

void insertLogElem(int b) {
   if (next < buffer_size) {
      buffer[next] = b;
      next = (next+1)%buffer_size;
      assert(next<buffer_size);
   }
}
```

**Figure 4. Implementation of a circular buffer.**

The pulse oximeter sources contain seven test cases, which *intend* to cover all possible execution paths related to the circular buffer, and during dynamic verification, we are not able to find any bug in the circular buffer implementation with these. However, the implementation is flawed: the array *buffer* is declared to be of type *char*[] (see line 1 in Figure 4) but we assign an element *b* of type *int* (see line 14) The test cases do not uncover this error because they happen to use only integer values that can safely be cast to a *char*.

Using SMT-based BMC, we can detect this bug by non-detemininistically assigning a value to the parameter *b* (i.e., by adding an assignment *b* = *nondet_int()* after

```
static void testCircularBuffer(void){
   int senData[]={1, -128, 98, 88, 59,
                  1, -128, 90, 0, -37};
   int i;
   initLog(5);
   for(i=0; i<10; i++)
      insertLogElem(senData[i]);
   for(i=5; i<10; i++)
      TEST_ASSERT_EQUAL_INT(senData[i],
                            removeLogElem());
}
```

**Figure 5. A unit test for the functions shown in Figure 4.**

line 12 in Figure 4). However, in general this approach can lead to false positives because the non-deterministic choice of values for program variables may force the exploration of paths that are infeasible in the original program. Rather than modifying the program we thus modify the test stubs and replace the concrete input values by non-deterministic choices. Here, we replace the initialization of the array *senData* (see line 2 and 3 of Figure 5) by *int senData*[] = {*nondet_int*(),...,*nondet_int*()}. We then use *assume*-statements to force the model checker away from the values that have already been explored during testing. In order to block larger parts of the search space, we use the given concrete values from all stubs and combine the respective values into a single interval for each variable or array element; here we assume that all "obvious" boundary values are used in some of the stubs, so that we are not blocking parts of the search space contain "obvious" errors but force the model checker towards the "unobvious" errors. In the example, we thus add an *assume*-statement such as *assume*(*senData*[0]<=1 && *senData*[0]>=42) and are then able to find two bugs related to overflow and underflow.

## 3 SMT-based Bounded Model Checking of Embedded ANSI-C Software

In BMC, the program to be analyzed is modelled as a state transition system, which is built by extracting its behaviour from the control-flow graph (CFG) [17]. This graph is used as part of a translation process from program text to SSA-form. A node in the CFG represents either a deterministic or non-deterministic assignment or a conditional statement, while an edge in the CFG represents the ability of the program to change the control location.

A state transition system $M = (S, S_0, \gamma)$ is an abstract machine that consists of a set of states $S$ (where $S_0 \subseteq S$

represents the set of initial states) and $\gamma \subseteq S \times S$ is the concrete transition relation [5]. A state $s \in S$ consists of the value of the program counter *pc* and the values of all program variables. An initial state $s_0$ assigns the initial program location of the CFG to the *pc*. We identify each transition $\gamma = (s_i, s_{i+1})$ between two states $s_i$ and $s_{i+1}$ with a logical formula $\gamma(s_i, s_{i+1})$ that captures the constraints on the corresponding values of the program counter and the program variables.

Given a transition system *M*, a property $\phi$, and a bound *k*, BMC unrolls the system *k* times and translates it into a verification condition $\psi$ such that $\psi$ is satisfiable if and only if $\phi$ has a counterexample of depth less than or equal to *k*. The VC $\psi$ is a quantifier-free formula in a decidable subset of first-order logic, which is then checked for satisfiability by an SMT solver. In this work, we are interested in checking two classes of properties: safety and liveness. The model checking problem associated with SMT-based BMC for checking safety properties is then formulated by constructing the following logical formula [1, 12]:

$$\psi^k = \underbrace{I(s_0) \wedge \bigwedge_{i=0}^{k-1} \gamma(s_i, s_{i+1})}_{constraints} \wedge \overbrace{\neg \phi_G^k}^{property} \quad (2)$$

where $\phi_G^k$ represents a safety property $\phi_G$ in step *k*, *I* is the function for the set of initial states of *M* and $\gamma(s_i, s_{i+1})$ is the function of the transition relation of *M* at time steps *i* and *i* + 1. Hence, the formula $\bigwedge_{i=0}^{k-1} \gamma(s_i, s_{i+1})$ unrolls the transition system and then represents the set of all executions of *M* up to the length *k* or less. $\neg \phi_G^k$ represents the condition that $\phi_G$ is violated by a bounded execution of *M* of length *k* or less. A counterexample for a property $\phi_G$ is a sequence of states $s_0, s_1, \ldots, s_k$ with $s_0 \in S_0$, $s_k \in S$, and $\gamma(s_i, s_{i+1})$ for $0 \leq i < k$. Liveness properties $\phi_F$ are checked by encoding $\neg \phi_F^i$ in a loop within a bounded execution of length at most *k*, such that $\phi_F$ is negated on each state in the loop [3]. In this case, formula (2) is rewritten as:

$$\psi^k = I(s_0) \wedge \bigwedge_{i=0}^{k-1} \gamma(s_i, s_{i+1}) \wedge \left( \bigvee_{i=0}^{k} \neg \phi_F^i \right) \quad (3)$$

For further information about the main software components of our ESBMC tool, we refer the reader to [8].

### 3.1 Optimizations

ESBMC implements some standard code optimization techniques such as constant folding, forward substitution, and reduction of variables [17]. In our previous work [8], we described how constant folding and forward substitution are implemented in the context of bounded model checking.

```
int main() {
int a[2], i, x;
if (x==0)
   a[i]=0;
else
   a[i+2]=1;
assert(a[i+1]==1);
}
```
(a)

```
g1 == (x1 == 0)
a1 == (a0 WITH [i0:=0])
a2 == a0
a3 == (a2 WITH [2+i0:=1])
a4 == (g1 ? a1 : a3)
t1 == (a4[1+i0] == 1)
```
(b)

**Figure 6. (a) A C program with violated property. (b) The C program of (a) in SSA form.**

Here, we focus on how to reduce the number of redundant variables generated from the BMC instances.

We use the code in Figure 6 as a running example. Figure 6(a) shows a syntactically valid C program that accidentally writes to an address outside the allocated memory region of the array *a* (line 6). Figure 6(b) shows the program in SSA form, where it only consists of *if* instructions, assignments and assertions. As we can see in Figure 6(b), for each assignment of the form *a = expr* (lines 4 and 6 from Figure 6(a)), the left-hand side variable is replaced by a new variable (e.g., $a_1$, $a_3$). In addition, the *if* condition in line 3 is replaced by a guard (e.g., $g_1$), the array index *i* in lines 3, 6 and 7 is replaced by $i_0$ (note that the variable *i* is not modified in the C program of Figure 6(a)) and the *assert* macro in line 7 is replaced by a Boolean variable $t_1$. From this representation, we build the constraints (*C*) and properties (*P*) formulae shown in (4) and (5) using the quantifier-free fragment of the first-order theories (e.g., linear arithmetic, arrays), which are then passed to an SMT solver as $C \wedge \neg P$ to check satisfiability.

$$C := \begin{bmatrix} g_1 := (x_1 = 0) \\ \wedge \ a_1 := store(a_0, i_0, 0) \\ \wedge \ a_2 := a_0 \\ \wedge \ a_3 := store(a_2, 2 + i_0, 1) \\ \wedge \ a_4 := ite(g_1, a_1, a_3) \end{bmatrix} \quad (4)$$

$$P := \begin{bmatrix} i_0 \geq 0 \wedge i_0 < 2 \\ \wedge \ 2 + i_0 \geq 0 \wedge 2 + i_0 < 2 \\ \wedge \ 1 + i_0 \geq 0 \wedge 1 + i_0 < 2 \\ \wedge \ select(a_4, i_0 + 1) = 1 \end{bmatrix} \quad (5)$$

However, the SSA transformation in the CBMC frontend introduces redundant variables, which creep into the formula passed into the SMT-solver, thus increasing the search space. Figure 6(b) shows an example of an additional (redundant) variable (i.e., $a_2$) created in order to keep the value of array $a$ before the *if* statement in line 3 of Figure 6(a). In our back-end, we eliminate functionally determined variables by substitution. In our running example, $a_2 == a_0$ holds in all states, so we can eliminate this constraint and the variable $a_2$ by substituting every occurrence of $a_2$ by $a_0$. we thus rewrite (4) as:

$$C := \begin{bmatrix} g_1 := (x_1 = 0) \\ \land\, a_1 := store(a_0, i_0, 0) \\ \land\, a_3 := store(a_0, 2 + i_0, 1) \\ \land\, a_4 := ite(g_1, a_1, a_3) \end{bmatrix} \quad (6)$$

Eliminating functionally determined variables allows us to remove most of the intermediate variables generated during the symbolic execution and consequently the size of the final formula to be sent to the SMT solver.

## 3.2 Encodings

In this section, we focus on new encodings (compared to our previous work [8]) that allow us to analyze more complex ANSI-C programs and show how to exploit different data types and solvers to speed up the verification process.

### 3.2.1 Fixed-Point Arithmetic

Embedded applications from discrete control and telecommunications domains often require arithmetic on non-integral numbers. However, full floating-point arithmetic is too heavyweight to be encoded into the BMC framework; instead, we approximate it by fixed-point arithmetic and use two different representations: in base 2 (when dealing with bit-vector arithmetic) and in base 10 (when dealing with rational arithmetic). Hence, in fixed-point arithmetic, we encode the numbers using the integer part and fractional part (i.e., after the radix point ".") [16]. For instance, the numbers 0.75 and 0.125 are fixed-point numbers with two-digit and three-digit fractional parts respectively. The number 0.75 can be represented as $\langle 0000.11 \rangle$ in base 2 or 3/4 in base 10 while the number 0.125 can represented as $\langle 0000.0010 \rangle$ in base 2 or 125/1000 in base 10. Given a rational number $R$ that consists of an integer part $I$ with $m$ bits and a fraction part $F$ with $n$ bits, we represent $R$ by $I.F$ and denote it by $\langle I.F \rangle / 2^n$.

Using bit-vector arithmetic, we encode fixed-point arithmetic exactly as in binary encoding and assume that the numbers have the same bitwidth before and after the radix point. If the numbers do not have the same bitwidth (e.g., if we want to sum 0.75 and 0.125), we add zeros from the right in the fractional part if there are bits missing after the radix point (e.g., $0.75 + 0.125 = \langle 0000.1100 \rangle + \langle 0000.0010 \rangle$) or we add bits from the left using sign-extension if there are bits missing before the radix point. On the other hand, using rational arithmetic, we encode fixed-point numbers by mapping $\{0, 1\}^{m+n} \to \mathbb{Q}$, which extracts the integer part and divides the fraction part $\langle F \rangle / 2^n$ and finally converts the resulting number to a rational one in base 10. As a result, the arithmetic operations are performed in the domain of $\mathbb{Q}$ instead of $\mathbb{R}$ and there is no need to add missing bits to the integer and fractional parts. In general, the drawback is that some numbers are not precisely represented with fixed-point arithmetic. However, we have not detected any false positives due to use of fixed-point arithmetic in any of our benchmarks.

### 3.2.2 Dynamic Memory Allocation

Although dynamic memory allocation is discouraged in embedded software, ESBMC is capable of model checking programs that use it through the ANSI-C functions *malloc* and *free*. ESBMC checks two properties related to *dynamic memory allocation*: *(i)* whether the argument to any *malloc*, *free*, or dereferencing operation is a dynamic object (*IS_DYNAMIC_OBJECT*) and *(ii)* whether the argument to any *free* or dereferencing operation is still a valid object (*VALID_OBJECT*).

Formally, let $p_o$ be a pointer expression that points to the object $o$ of type $t$ and let $a$ be an array of type $t$ and size $n$, where $n$ is of integer type and represents the number of elements to be allocated. In our encoding, each dynamic object has an unique identifier denoted by $\rho_k$, where the subscript $k$ indicates the objects "serial number" in sequential order of all dynamically created objects. Let $i$ be an integer variable that indicates the position in which the object pointed to by $p_o$ must be stored in array $a$. We thus encode *IS_DYNAMIC_OBJECT* as a literal $l_d$ with the following constraint:

$$l_d \Leftrightarrow \left( \bigvee_{n=0}^{k-1} \rho_n = p_o \right) \land (0 \leq i < n) \quad (7)$$

To check for invalid objects, we add one additional bit field $\nu$ to each dynamic object to indicate whether it is still alive or not. Let $\bot$ denote the state that the object is not alive. We then encode *VALID_OBJECT* as a literal $l_v$ with the following constraint:

$$l_v \Leftrightarrow (p_o.\nu \neq \bot) \quad (8)$$

### 3.2.3 Exploiting Datatypes and Solvers

Modern SMT solvers provide ways to model the program variables as bit-vectors or as elements of a numerical do-main (e.g., $\mathbb{Z}$, $\mathbb{Q}$, or $\mathbb{R}$). If the program variables are modelled as bit-vectors of fixed size, then the result of the analysis can be precise (w.r.t. the ANSI-C semantics), depending on the size considered for the bit-vectors. On the other hand, if the program variables are modelled as numerical values, then the result of the analysis is independent from the actual binary representation, but the analysis may not be precise when arithmetic expressions are involved. As a motivating example, consider the following small C program from [6] as shown in Figure 7.

```c
int main() {
  unsigned char a, b;
  unsigned int result=0, i;
  a=nondet_uchar();
  b=nondet_uchar();
  for(i=0; i<8; i++)
    if((b>>i)&1)
      result+=(a<<i);
  assert(result==a*b);
}
```

**Figure 7. A C program that uses shift-and-add to multiply two numbers.**

This program nondeterministically selects two values of type *unsigned char* and uses bitwise AND, right- and left-shift operations to multiply them. Reasoning about this program by means of integer arithmetic produces wrong results if the bit-level operators are treated as uninterpreted functions (UFs). Although UFs simplify the proofs, they ignore the semantics of the operators and consequently make the formula weaker. In addition, the majority of the software model checkers (e.g., SMT-CBMC [1] and BLAST [14]) fail to check the assertion in line 9. On the other hand, bit-vector arithmetic allows us to encode bit-level operators in a more accurate way. However, in our benchmarks, we noted that the majority of VCs are solved faster if we model the basic datatypes as integer and/or real. Consequently, we have to trade off between *speed* and *accuracy* which might be two competing goals in formal verification using SMT.

Based on the extent to which the SMT solvers support the domain theories and on experimental results obtained with a large set of benchmarks, we developed a simple but effective heuristic to determine the best representation for the program variables as well as the best SMT-solver to be used in order to check the properties of a given ANSI-C program. Our default representation for encoding the constraints and properties of ANSI-C programs are integers and reals, respectively, and our default solver is Z3. We then explore the CFG representation of the program. If we find expressions that involve bit operations (e.g., $<<$, $>>$, $\&$, $|$, $\oplus$) or typecasts from signed to unsigned datatypes and vice-versa, we encode the corresponding variables as bit-vectors and switch the SMT solver either to Boolector (if no pointers are used) or Z3 (if pointers are used).

## 4 Experimental Evaluation

The experimental evaluation of our work consists of three parts. Section 4.1 contains the results of applying ESBMC to the verification of six sets of ANSI-C benchmarks. The purpose of this first part is to check the error-detection capability of ESBMC since most of these benchmarks contain ANSI-C programs with and without bugs. Section 4.2 contains the experimental results of applying ESBMC and CBMC to the verification of embedded software used in the telecommunications domain. The purpose of this second part is to evaluate ESBMC's relative performance against CBMC using embedded software industrial applications. Section 4.3 contains the results of applying the continuous verification approach, described in Section 2, to large embedded software used in a commercial product.

All experiments were conducted on an otherwise idle Intel Pentium Dual CPU, 2GHz with 4 GB of RAM running Linux OS. For all benchmarks, the time limit has been set to 3600 seconds for each individual property. All times given are wall clock time in seconds as measured by the unix `time` command through a single execution.

### 4.1 Error-Detection Capability

As a first step, we analyze to which extent ESBMC is able to handle standard ANSI-C benchmarks. For this purpose, we analyzed the programs in the VERISEC, NECLA, SNU-RT, PowerStone, and WCET benchmarks. VERISEC and NECLA are not related to embedded software, but they allow us to check the error-detection capability of the model checkers since these benchmarks provide ANSI-C programs with and without bugs including dynamic memory allocation, interprocedural dataflow, aliasing, and string manipulation. The remaining three benchmarks contain ANSI-C constructs found in embedded software.

Table 1 summarizes the results of applying ESBMC to the verification of the standard ANSI-C benchmarks; a detailed description of these experimental results is available at users.ecs.soton.ac.uk/lcc08r/esbmc. Here, #L denotes the total number of lines of code, B denotes the unwinding bound, and #P denotes the number of properties to be verified for each ANSI-C program. The *Time* columns contain the total time in seconds to check all properties of a given ANSI-C program, broken down into frontend (i.e., encoding) and backend (i.e., decision procedure). *Failed* indicates how many properties failed during the verification process; properties can fail for two reasons: either due to time out (TO) or due to memory out (MO).

|   | Module | #L | #P | Time | | | #P | | |
|---|--------|-----|------|----------|--------------------|---------|--------|----------|--------|
|   |        |    |    | Encoding | Decision Procedure | Total   | Passed | Violated | Failed |
| 1 | VERISEC | 9090 | 2148 | 190.85 | 228.35 | 419.2 | 1946 | 202 | 0 |
| 2 | NECLA | 1011 | 208 | 59.28 | 88.9 | 148.18 | 188 | 20 | 0 |
| 3 | SNU-RT | 3102 | 790 | 3166.94 | 12.18 | 3179.12 | 762 | 28 | 0 |
| 4 | WCET | 3430 | 726 | 72.72 | 8.28 | 81 | 722 | 4 | 0 |
| 5 | POWERSTONE | 2957 | 2053 | 127.36 | 913.64 | 1041 | 2043 | 10 | 0 |

**Table 1. Results of the error-detection capability with ESBMC.**

As mentioned previously, benchmarks VERISEC and NECLA contain ANSI-C programs with and without bugs and all inputs are replaced by nondeterministic one. In these benchmarks, ESBMC is able to detect common design errors related to buffer overflow, aliasing, dynamic memory allocation, and string manipulation. In the remaining benchmarks, ESBMC is able to model check ANSI-C programs that involve tight interplay between non-linear arithmetic, bit operations, pointers and array manipulations. In addition, ESBMC was able to find undiscovered bugs in the SNU-RT, WCET, and PowerStone benchmarks related to arithmetic overflow, invalid pointer and pointer arithmetic. We also checked the effect of eliminating functionally determined variables and we observed that the total verification time can only be reduced by 0.3% to 1% in the benchmarks. We believe that the SMT solvers already eliminate the redundant variables during the pre-processing phase.

## 4.2 Comparison to CBMC

In order to evaluate ESBMC's performance relative to CBMC, we analyzed the embedded software used in a commercial product from NXP semiconductors [18], a set-top box that is used in high definition internet protocol (IP) and hybrid digital TV applications. The embedded software of this platform relies on the Linux operating system and use the *LinuxDVB*, *DirectFB* and *ALSA* applications.

We evaluated CBMC version 3.3.2 and we invoked both tools (i.e., CBMC and ESBMC) by setting manually the file name, the unwinding bound and the overflow check. CBMC has support for SAT and SMT solvers in the back-end and in our comparison we use the SMT solver Z3 for evaluating both tools CBMC and ESBMC. Table 2 shows the results when applying SAT-based CBMC, SMT-based CBMC and ESBMC to the verification of the embedded applications. SAT-based CBMC performs better than SMT-based CBMC since it is able to analyze one additional embedded application (viz. exStbCc).

CBMC is not able to check the modules *exStbKey*, *exStbFb*, and *exStbDemo* due to memory limitations as well as modules *exStbResolution* and *exStbHDMI* due to time out and segmentation fault. Both CBMC and ESBMC fail to model check the module *exStbDemo*, which contains approximately 31KLOC and represents the largest module that we analyzed. However, the results indicate that ESBMC scales significantly better than CBMC for problems that involve tight interplay between non-linear arithmetic, bit operations, pointers and arrays.

## 4.3 Scalability

In order to model check the *exStbDemo* module we apply the continuous verification approach as described in Section 2; Table 3 summarizes the results. Here, #TC denotes the total number of test cases. For the dynamic verification, we use the EmbUnit[1] unit test framework, which provides means to apply assertion-based verification in embedded software written in C through a set of macros to assert strings, integers, pointer value, and conditions.

As described in Section 4.2, the state-of-the-art model checkers fail to verify the properties of the *exStbDemo* application due to memory limitations. However, if we use the test cases to guide the state space exploration, we can not only define the functions, but also the state variables that were not yet fully explored during dynamic verification. As a result, in the function *commandLoop*, ESBMC finds a property violation related to an invalid pointer in 4.39 seconds using an unwinding bound of 6. However, we are not able to increase further the unwinding bound of functions *commandLoop* and *getCommand* due to memory limitations.

We had access, from the NXP development team, to four different product releases (PRs) that contain the application *exStbDemo*. Based on these four PRs, we identified the functions and methods that have actually been modified and focus on these since the computational effort to re-verify each system build from scratch is too high. The four PRs are shown on the right-hand side in Table 3 as PR 10, 11, 12, 13. The development time of each PR is about one month and each one contains new features, enhancements (through *refactoring*), and bug fixes. We use PR10 as a reference (and starting point) to compare with PR11, PR11 to compare against PR12 and so on. The functions 2-8 shown in Table 3 do not present input/output relations and the function *getCommand* is not equivalent in PR10, PR11 and PR12. However, if we compare PR10 to PR11 and PR12, we can reduce the verification time by up to 50% since functions *checkEndOfIPStream* and *checkEndOfMediaStream* are not modified in four different versions. In addition, the function *commandLoop*, which represents one of the hardest functions, is only modified in PR13.

---

[1]EmbUnit. http://embunit.sourceforge.net/

|   | Module | #L | B | #P | SAT-based CBMC | | | | | SMT-based CBMC | | | | | ESBMC | | | | |
|---|--------|----|---|----|---|---|---|---|---|---|---|---|---|---|---|---|---|---|---|
|   |        |    |   |    | Time | | #P | | | Time | | #P | | | Time | | #P | | |
|   |        |    |   |    | Decision Procedure | Total | Passed | Violated | Failed | Decision Procedure | Total | Passed | Violated | Failed | Decision Procedure | Total | Passed | Violated | Failed |
| 1 | exStbKey | 558 | 1 | 22 | 0.1 | 3.13 | 22 | 0 | 0 | 0.18 | 3.13 | 22 | 0 | 0 | 0.02 | 1.09 | 22 | 0 | 0 |
| 2 | exStbHDMI | 1508 | 11 | 41 | † | † | 0 | 0 | 41 | † | † | 0 | 0 | 41 | 128.44 | 211.02 | 41 | 0 | 0 |
| 3 | exStbLED | 430 | 10 | 59 | MO | MO | 0 | 0 | 59 | TO | TO | 0 | 0 | 59 | 1781 | 1817.61 | 59 | 0 | 0 |
| 4 | exStbHwAcc | 1432 | 1000 | 115 | 0.146 | 2.44 | 115 | 0 | 0 | 0.71 | 3.05 | 115 | 0 | 0 | 0.013 | 0.937 | 115 | 0 | 0 |
| 5 | exStbResolution | 353 | 150 | 32 | TO | TO | 0 | 0 | 32 | TO | TO | 0 | 0 | 32 | 1179.34 | 1596.62 | 32 | 0 | 0 |
| 6 | exStbFb | 689 | 30 | 48 | MO | MO | 0 | 0 | 48 | † | † | 0 | 0 | 48 | 50.53 | 138.31 | 48 | 0 | 0 |
| 7 | exStbCc | 331 | 200 | 5 | 174.49 | 198.19 | 5 | 0 | 0 | TO | TO | 0 | 0 | 5 | 31.24 | 46.13 | 5 | 0 | 0 |
| 8 | exStbDemo | 30902 | 17 | 267 | MO | MO | 0 | 0 | 267 | MO | MO | 0 | 0 | 267 | MO | MO | 0 | 0 | 267 |

**Table 2. Results of the comparison between CBMC and ESBMC. Time-outs are represented with TO in the Time column; Examples that exceed available memory are represented with MO in the Time column; Internal errors in the tool are represented with † in the Time column.**

|   | Function | B | #P | ESBMC | | | | | EmbUnit | | Subversion | | | |
|---|----------|---|----|---|---|---|---|---|---|---|---|---|---|---|
|   |          |   |    | Time (s) | | #P | | | | | | | | |
|   |          |   |    | Decision Procedure | Total | Passed | Violated | Failed | #TC | Time (s) | PR10 | PR11 | PR12 | PR13 |
| 1 | getCommand | 6 | 237 | <0.2 | 4.39 | 236 | 1 | 0 | 18 | <0.1 | X | X | X | |
| 2 | commandLoop | 6 | 237 | 70.22 | 128.41 | 237 | 0 | 0 | 26 | <0.1 | X | | | X |
| 3 | checkCommandParams | 17 | 229 | 73.29 | 161.14 | 229 | 0 | 0 | 4 | <0.1 | X | X | X | X |
| 4 | checkEndOfPvrStream | 20 | 228 | <0.1 | 4.34 | 228 | 0 | 0 | 3 | <0.1 | X | | | X |
| 5 | checkEndOfIPStream | 20 | 228 | <0.1 | 3.99 | 228 | 0 | 0 | 3 | <0.1 | X | | | |
| 6 | checkEndOfMediaStream | 20 | 228 | <0.1 | 3.97 | 228 | 0 | 0 | 4 | <0.1 | X | | | |
| 7 | setupFramebuffers | 17 | 228 | <0.1 | 4.22 | 228 | 0 | 0 | 2 | <0.1 | X | X | | X |
| 8 | setupFBResolution | 17 | 228 | <0.1 | 3.92 | 228 | 0 | 0 | 2 | <0.1 | X | | X | |
| | Total verification time in seconds for each PR | | | | | | | | | | 314.38 | 169.75 | 169.45 | 298.11 |

**Table 3. Results of the continuous verification approach.**

## 5 Related Work

One way of tackling large verification problems is to leverage both parallelism and search diversity [15]. Holzmann et al. describe the Swarm tool that allows to use different search strategies on multi-core machines [15]. It is the main interface to the SPIN model checker to verify larger systems. This approach, however, involves large communication overhead and does not take into account information from the SCM system in order to focus the verification effort on new and/or modified functions.

Peled proposes a set of combinations between model checking and testing, which includes black box checking, adaptive model checking, and unit checking [19]. However, he does not consider the development history from the SCM system and also uses explicit model checking based on automata theory, which does not scale well due to the number of program variables and data type widths [10]. In addition, Peled only describes the techniques, but does not apply it to any commercial product. Gunter and Peled [13] extend this approach by proposing a symbolic verification approach for a unit of code, also called unit checking. The authors, however, apply this approach only to check whether a complex number diverges to infinity, while we focus on the verification of large embedded software. Sen proposes an approach to execute a program concretely and symbolically by combining random testing and symbolic execution [20]. This approach, however, might fail to compute concrete values that satisfy a given path constraint due to the constraint solver performance.

SMT-based BMC is gaining popularity in the formal verification community due to the advent of sophisticated SMT solvers built over efficient SAT solvers [4, 9]. Ganai and Gupta describe a verification framework for BMC which extracts high-level design information from an extended finite state machine (EFSM) and applies several techniques to simplify the BMC problem [12]. However, the authors flatten the structures and arrays into scalar variables in such a way that they use only the theory of integer and real arithmetic, which does not reflect precisely the ANSI-C semantics. Armando et al. also propose a BMC approach using SMT solvers for ANSI-C programs [1]. In this approach, however, they only make use of linear arithmetic (addition and multiplication by constants), arrays, records and bit-vectors and as a consequence, their SMT-CBMC prototype does not address important constructs of the ANSI-C language (e.g., non-linear arithmetic and bit-shift operations).

Recently, a number of static checkers have been developed in order to trade off scalability and precision. Calysto is a static checker that is able to verify VCs related to arithmetic overflow, nil-pointer dereferencing and assertions specified by the user [2]. The VCs are passed to the SMT solver SPEAR which supports bit-vector arithmetic and is customized for the VCs generated by Calysto. However, Calysto does not support float-point operations and unsoundly approximates loops by unrolling them only once. As a consequence, soundness is relinquished for performance. Saturn is another efficient static checker that scales to larger systems, but with the drawback of losing precision by supporting only the most common integer operators and performing at most two unwindings of each loop [22].

## 6 Conclusions

In this work, we have defined a new concept called *continuous verification* and we have applied it to the verification of large embedded software used in the telecommunications domain. In addition, we have described a new set of encodings that allow us to reason accurately about embedded software by discovering subtle bugs in several industrial applications and to scale for large embedded software. Furthermore, the experimental results show that our SMT-based model checker (ESBMC) outperforms both the SAT-based and SMT-based CBMC [6] model checker if we consider the verification of embedded software. As a result, our approach represents a promising direction to improve the state space coverage and to verify quickly the negation of properties in larger state spaces.

## References

[1] A. Armando, J. Mantovani, and L. Platania. Bounded model checking of software using SMT solvers instead of SAT solvers. *Int. J. Softw. Tools Technol. Transf.*, vol(11), pp. 69–83, 2009.

[2] D. Babić and A. J. Hu. Calysto: Scalable and Precise Extended Static Checking. In *ICSE*, pp. 211–220, 2008.

[3] A. Biere. Bounded model checking. In *Handbook of Satisfiability*, pp. 457–481. 2009.

[4] R. Brummayer and A. Biere. Boolector: An efficient SMT solver for bit-vectors and arrays. In *TACAS*, LNCS 5505, pp. 174–177, 2009.

[5] E. Clarke, O. Grumberg, and D. Peled. *Model Checking*. The MIT Press, Cambridge, MA, 2000.

[6] E. Clarke et al. A tool for checking ANSI-C programs. In *TACAS*, LNCS 2988, pp. 168–176, 2004.

[7] L. Cordeiro et al. Agile development methodology for embedded systems: A platform-based design approach. In *ECBS*, pp. 195–202, 2007.

[8] L. Cordeiro, B. Fischer, and J. Marques-Silva. SMT-based bounded model checking for embedded ANSI-C software. To appear in ASE. http://eprints.ecs.soton.ac.uk/18166/. 2009.

[9] L. M. de Moura and N. Bjørner. Z3: An efficient SMT solver. In *TACAS*, LNCS 4963, pp. 337–340, 2008.

[10] V. D'Silva, D. Kroening, and G. Weissenbacher. A survey of automated techniques for formal software verification. *IEEE Trans. on CAD of Integrated Circuits and Systems*, 27(7):1165–1178, 2008.

[11] M. Fowler. *Continuous Integration*. ThoughtWorks. http://martinfowler.com, 2006.

[12] M. K. Ganai and A. Gupta. Accelerating high-level bounded model checking. In *ICCAD*, pp. 794–801, 2006.

[13] E. L. Gunter and D. Peled. Model checking, testing and verification working together. *Formal Asp. Comput.*, 17(2):201–221, 2005.

[14] T. A. Henzinger et al. *BLAST: Berkeley Lazy Abstraction Software Verification Tool*. http://mtc.epfl.ch/software-tools/blast/, 2009.

[15] G. J. Holzmann, R. Joshi, and A. Groce. Tackling large verification problems with the swarm tool. In *SPIN*, LNCS 5156, pp. 134–143, 2008.

[16] D. Kroening and O. Strichman. *Decision Procedures: An Algorithmic Point of View*. Springer Publishing Company, Incorporated, 2008.

[17] S. S. Muchnick. *Advanced compiler design and implementation*. Morgan Kaufmann Publishers Inc., 1997.

[18] NXP. *High definition IP and hybrid DTV set-top box STB225*. http://www.nxp.com/, 2009.

[19] D. Peled. Model checking and testing combined. In *ICALP*, pp. 47–63, 2003.

[20] K. Sen. Concolic testing. In *ASE*, pp. 571–572, 2007.

[21] F. Somenzi and R. Bloem. Efficient Büchi automata from LTL formulae. In *In CAV*, LNCS 1855, pp. 247–263, 2000.

[22] Y. Xie and A. Aiken. Scalable error detection using Boolean satisfiability. *SIGPLAN Not.*, vol(40), pp. 351–363, 2005.